%% file: main.tex
\documentclass[10pt, conference, compsocconf]{IEEEtran}
\IEEEoverridecommandlockouts
\usepackage[utf8]{inputenc}
\usepackage[T1]{fontenc}
\usepackage[english]{babel}
\usepackage{ntheorem}
\usepackage{amsmath,amssymb,amstext}
\usepackage{epsfig,psfrag}
\usepackage{graphics,graphicx} 
\usepackage{color}
\usepackage{dsfont}
\usepackage{url}
\usepackage{alltt}
\usepackage{stmaryrd}
\usepackage{color}
\usepackage[font=footnotesize]{subfig}
\usepackage{array,algorithm2e,algorithmic}
\renewcommand{\theequation}{\arabic{equation}}
\usepackage[width=18.2cm,height=25.7cm]{geometry}
\usepackage{epsfig}

\newtheorem{definition}{Definition}

\newtheorem{proposition}{Proposition}

\newtheorem{theorem}{Theorem}

\SetKwInput{KwData}{Parameters}

\input{macroE}

\begin{document}

\title{Performance Analysis of a Keyed Hash Function based
  on Discrete and Chaotic Proven Iterations}

\author{Jacques M. Bahi, Jean-Fran\c{c}ois Couchot, and Christophe Guyeux*\thanks{* Authors in alphabetic order}\\
University of Franche-Comté, Computer Science Laboratory (LIFC)\\
Belfort, France\\
Email: \{jacques.bahi, jean-francois.couchot, christophe.guyeux\}@univ-fcomte.fr}

\maketitle

\begin{abstract}
\input{abstract}
\end{abstract}


\section{Introduction}\label{sec:intro}
\input{intro}

\section{Discrete and Chaotic Proven Iterations}\label{sec:chaos}
\input{chaos}

\section{A Chaos-Based Keyed Hash Function}\label{sec:algo}
\input{hash}

\section{Qualitative Analysis}\label{sec:eval}

\input{avalanche}

\input{preimage}

\input{confusion}

\section{Quantitative and Experimental Evaluations}

\input{performance}

\input{perf}

\section{Conclusion}
\input{conclusion}

\bibliographystyle{IEEEtran}
\bibliography{mabase,mabase2}

\end{document}

%% file: macroE.tex
\newcommand{\Nats}[0]{\ensuremath{\mathds{N}}}

\newcommand{\R}[0]{\ensuremath{\mathds{R}}}
\newcommand{\Bool}[0]{\ensuremath{\mathds{B}}}



%% file: abstract.tex
Security of information transmitted through the Internet is an international concern.
This security is guaranteed by tools like hash functions.
However, as security flaws have been recently identified in the current standard in this domain, new ways to hash digital media must be investigated.
In this document an original keyed hash function is evaluated.
It is based on chaotic iterations and thus possesses various topological properties as uniform repartition and sensibility to its initial condition.
These properties make our hash function satisfy the requirements in this field.
This claim is verified qualitatively and experimentally in this research work, among other things by realizing simulations of diffusion and confusion.

%% file: intro.tex
Hash  functions are  fundamental tools  to guarantee  the  quality and
security of  data exchanges through the Internet.   For instance, they
allow to  store passwords  in a  secure manner or  to check  whether a
download has occurred  without any error.  SHA-1 is  probably the most
widely  used  hash functions.   It  is present  in  a  large panel  of
security applications and protocols through the Internet.  However, in
the last decade,  security flaws have been detected  in SHA-1.  As the
SHA-2 variants are algorithmically  close to SHA-1 and produce finally
message  digests on  principles similar  to  the MD4  and MD5  message
digest algorithms, a new hash standard based on original approaches is
then eagerly  awaited.  In this context,  we have proposed  a new hash
function  in~\cite{guyeux09}.   Based on  chaotic iterations,
this  function behaves completely  different from  approaches followed
until now.

However  chaos  insertion  to  produce  hash  functions  is  sometimes
disputed~\cite{Zhou1997429,Guo20093201}.   Indeed existing chaos-based
hash functions  only include  ``somewhere'' some chaotic  functions of
real variables like logistic, tent,  or Arnold's cat maps.  It is 
then supposed  that  the final  hash  function  preserves these  
properties~\cite{Wang2003,Xiao20094346,Xiao20092288,Xiao20102254}.   But,  in our
opinion,  this claim  is  not  so evident.   Moreover,  even if  these
algorithms are themselves proven  to be chaotic, their implementations
on finite machines can result  to lost of chaos property.  Among other
things, the main  reason is that chaotic functions  (embedded in these
researches)  only manipulate  real numbers,  which do  not exist  in a
computer.   In~\cite{guyeux09}, the  hash  function we  have
proposed does not simply integrate chaotic maps into algorithms hoping
that the  result remains chaotic;  we have conceived an  algorithm and
have mathematically proven that it  is chaotic.  To do both our theory
and our implementation are based on finite integer domains and chaotic
iterations.

Chaotic iterations (CIs) were  formerly a way to formalize distributed
algorithms  through  mathematical  tools~\cite{Chazan69}.   Thanks  to
these  CIs,  it  was  thus   possible  to  study  the  convergence  of
synchronous or asynchronous  programs over parallel, distributed, P2P,
grid,  or GPU  platforms, in  a view  to solve  linear  and non-linear
systems.  CIs  have recently revealed  numerous interesting properties
of  disorder  formalized  into  the mathematical  topology  framework.
These studies  lead to the  conclusion that the  chaos of CIs  is very
intense and that chaos class  can tackle the computer science security
field~\cite{GuyeuxThese10}.  As  CIs only manipulate  binary digits or
integers,  we have  shown  that  they are  amenable  to produce  truly
chaotic  computer programs.   Among other  things, CIs
have       been       applied       to      pseudo-random       number
generators~\cite{wbg10:ip}  and   to  an  information  hiding
scheme~\cite{bg10:ip}  in the  previous sessions of  the International
Conference  on  Evolving  Internet.    In  this  paper,  the  complete
unpredictable behavior of chaotic iterations is capitalized to produce
a truly chaotic keyed hash function.

The remainder of this research work is organized in the following way.
In   Section~\ref{sec:chaos},   basic   recalls  concerning   chaotic
iterations and Devaney's chaos  are recalled.  Our keyed hash function
is  presented, reformulated, and  improved in  Section~\ref{sec:algo}.
Performance analyses  are presented in  the next two sections:  in the
first  one a  qualitative  evaluation of  this  function is  outlined,
whereas  in  the second  one  it  is  evaluated experimentally.   This
research work ends by a  conclusion section, where our contribution is
summarized and intended future work is given.

%% file: chaos.tex
This  section gives  some recalls  on topological  chaotic iterations.
Let us firstly discuss about  domain of iterated functions.  As far as
we know, no result rules that  the chaotic behavior of a function that
has  been   theoretically  proven  on   $\R$  remains  valid   on  the
floating-point numbers, which is  the implementation domain.  Thus, to
avoid loss of chaos this research work presents an alternative, namely
to iterate  boolean maps: results  that are theoretically  obtained in
that domain are preserved during implementations.

Let us denote by $\llbracket  a ; b \rrbracket$ the following interval
of  integers:   $\{a,  a+1,  \hdots,  b\}$.   A   {\em  system}  under
consideration  iteratively modifies  a  collection of  $n$~components.
Each  component $i  \in \llbracket  1; n  \rrbracket$ takes  its value
$x_i$ among the domain  $\Bool=\{0,1\}$.  A~{\em configuration} of the
system at discrete time $t$ (also  said at {\em iteration} $t$) is the
vector  $x^{t}= (x_1^{t},  \ldots, x_{n}^{t})  \in  \mathds{B}^n$.  In
what follows,  the dynamics of  the system is particularized  with the
negation function $\neg :  \Bool^n \rightarrow \mathds{B}^n$ such that
$\neg(x)    =   (\overline{x_i},   \ldots,    \overline{x_n})$   where
$\overline{x_i}$ is the negation of $x_i$.


In the  sequel, the  {\em strategy} $S=(S^{t})^{t  \in \Nats}$  is the
sequence defining which  component is updated at time  $t$ and $S^{t}$
denotes its $t-$th term.  We introduce the function $F_{\neg}$ that is
defined for the negation function by:
$$\begin{array}{ccl}
F_{\neg}:  \llbracket1;n\rrbracket\times \mathds{B}^{n}
& \rightarrow  & 
\mathds{B}^{n} \\  
F_{\neg}(s,x)_j & =  & 

\left\{
\begin{array}{l}
\overline{x_j} \textrm{ if } j= s \\ 
x_{j} \textrm{ otherwise.} \enspace 
\end{array}
\right. 
\end{array}$$

\noindent With such a notation, configurations are defined for times 
$t=0,1,2,\ldots$ by:
\begin{equation}\label{eq:sync}   \left\{\begin{array}{l}   x^{0}\in
\mathds{B}^{n} \textrm{ and}\\
 x^{t+1}= F_{\neg}(S^t,x^{t}) \enspace .
\end{array} \right.
\end{equation}
\noindent Finally, iterations defined in~(\ref{eq:sync}), called ``chaotic iterations''~\cite{Chazan69}, can be described by
the following system
\begin{equation} 
\left\{
\begin{array}{lll} 
X^{0} & =&  ((S^t)^{t \in \Nats},x^0) \in 
\llbracket1;n\rrbracket^{\Nats}\times\mathds{B}^{n}\\ 
X^{k+1}& =& G_{\neg}(X^{k})
\end{array}
\right.
\enspace ,
\label{eq:Gf}
\end{equation}
\noindent such that  
$$
G_{\neg}\left(((S^t)^{t \in \Nats},x)\right) 
= \left(\sigma((S^t)^{t \in \Nats}),F_{\neg}(S^0,x)\right) 
$$

\noindent where  $\sigma$ is the  function that  removes the  first term  of the
strategy (\textit{i.e.},~$S^0$). 
Let us remark that the term ``chaotic'' in the name of this tool is just an adjective, which has a priori no link with the mathematical theory of chaos. 

In the space 
$\mathcal{X} = \llbracket 1 ; n \rrbracket^{\Nats} \times 
\Bool^n$ we define the distance between 
two points $X = (S,E), Y = (\check{S},\check{E})\in \mathcal{X}$ by%
\begin{eqnarray*}
d(X,Y)& =& d_{e}(E,\check{E})+d_{s}(S,\check{S}), \textrm{ where} \\
\displaystyle{d_{e}(E,\check{E})} & = & \displaystyle{\sum_{k=1}^{n%
}\delta (E_{k},\check{E}_{k})}, \textrm{ and} \\
\displaystyle{d_{s}(S,\check{S})} & = & \displaystyle{\dfrac{9}{n}%
\sum_{k=1}^{\infty }\dfrac{|S^k-\check{S}^k|}{10^{k}}}.%
\end{eqnarray*}

If the floor value $\lfloor d(X,Y)\rfloor $ is equal to $j$,
then the systems $E, \check{E}$ differ in $j$ cells. 
In addition, $d(X,Y) - \lfloor d(X,Y) \rfloor $ is a measure of the differences between strategies $S$ and $\check{S}$. More precisely, this floating part is less than $10^{-k}$ if and only if the first $k$
terms of the two strategies are equal. Moreover, if the $k^{th}$ digit is nonzero, then the $k^{th}$ terms of the two
strategies are different.

With this material it has been already proven that~\cite{GuyeuxThese10}:
\begin{itemize}
\item $G_{\neg}$ is a continuous function on a suitable metric space $(\mathcal{X},d)$, 
\item iterations as defined in Equ.~\ref{eq:Gf} 
are regular (\textit{i.e.}, periodic points of $G_{\neg}$ are dense in 
$\mathcal{X}$),
\item $(\mathcal{X},G_{\neg})$ is topologically transitive
(\textit{i.e.}, for any pair of open sets $U,V\subset \mathcal{X}$,
there exists some natural number $k>0$  s. t. 
$G_{\neg}^{k}(U)\cap V\neq \varnothing $),
\item $(\mathcal{X},G_{\neg})$ has sensitive dependence on initial conditions
(\textit{i.e.}, there exists $\delta >0$ s.t. for any $X\in \mathcal{X}$
and any neighborhood $V$ of $X$, there exist $Y\in V$ and $k\geqslant 0$
with $d(G_{\neg}^{k}(X), G_{\neg}^{k}(Y))>\delta $).
\end{itemize}

To sum up, we have previously established that the three conditions for Devaney's chaos hold for chaotic iterations.
So CIs behave chaotically, as it is defined in the mathematical theory of 
chaos~\cite{devaney,Knudsen94}.

%% file: hash.tex
This section first recalls an informal definition~\cite{BSP96,ZWZ07}
of Secure Keyed One-Way Hash Function. We next present our algorithm. Finally, 
we establish relations between  the algorithm properties inherited from topological results and requirements of Secure Keyed One-Way Hash
Function. 

\subsection{Secure Keyed One-Way Hash Function}

\begin{definition}[Secure Keyed One-Way Hash Function]
Let $\Gamma$ and $\Sigma$ be two alphabets,   
let $k \in K$ be a key in a given key space,
let $l$ be a natural numbers which is the length of the output message,
and let $h : K \times  \Gamma^{+} \rightarrow \Sigma^{l}$ be a function that associates 
a message in $\Sigma^{l}$ for each pair of key, word in  
$K \times  \Gamma^{+}$.
The set of all functions $h$ is partitioned into classes
of functions $\{h_k : k \in K \}$
indexed by a key $k$ and such that 
$h_{k}: \Gamma^{+} \rightarrow \Sigma^{l}$ is defined by
$h_{k}(m) = h(k,m)$  \textit{i.e.}, $h_{k}$ generates a message digest of length $l$.

A class $\{h_k : k \in K \}$ is a \emph{Secure Keyed One-Way Hash Function}  
if it satisfies the following properties:
\begin{enumerate}
\item the function $h_k$ is keyed one-way. That is,
  \begin{enumerate}
  \item Given $k$ and $m$, it is easy to compute $h_k(m)$ .
  \item Without knowledge of $k$, it is hard to find $m$ when $h_k(m)$ is given and to  find $h_k(m)$ when only $m$ is given.
  \end{enumerate}
\item The function $h_k$ is keyed collision free, that is, 
  without the knowledge of $k$ it is difficult to find two distinct messages
  $m$ and $m'$ s.t. $h_k(m)= h_k(m')$.
\item Images of function $h_k$ has to be uniformly distributed in 
$\Sigma^{l}$ in order to counter statistical attacks. 
\item Length $l$ of produced image has to be larger than $128$ bits 
 in order to counter birthday attacks.
\item Key space size has to be sufficiently large 
in order to counter exhaustive key search.
\end{enumerate}
\end{definition}    

Let us now present our hash function that is 
based on  chaotic iterations as defined in Section~\ref{sec:chaos}.
The  hash value message is obtained as the last configuration 
resulting from chaotic iterations of $G_{\neg}$.

We then have to define the pair $X^0=((S^t)^{t \in \Nats},x^0)$, \textit{i.e.},
the strategy and the initial configuration $x^0$.

\subsection{Computing $x^0$}
\label{subsec:computing x0}
The first step of the algorithm is to transform the message in a normalized
$n = 256$ bits sequence $x^0$.
This size $n$ of the digest can be changed, mutatis mutandis, if needed.
Here, this first step is close to the pre-treatment 
of the SHA-1 hash function, but it can easily be replaced by any other compression method.   

To illustrate this step, we take an example, our
original text is: ``\emph{The original text}''.

Each character of this string is replaced by its ASCII code (on 7 bits).
Following the SHA-1 algorithm, 
first we append a ``1'' to this string, which is then 

\begin{center}
\small
\begin{alltt}
 10101001 10100011 00101010 00001101 11111100
 10110100 11100111 11010011 10111011 00001110
 11000100 00011101 00110010 11111000 11101001.
\end{alltt}
\end{center}

Next we append the block 1111000, which is the binary value of this  
string length (120), and finally  another ``1'' is added:

\begin{center}
\small
\begin{alltt}
 10101001 10100011 00101010 00001101 11111100
 10110100 11100111 11010011 10111011 00001110
 11000100 00011101 00110010 11111000 11101001
 11110001.
\end{alltt}
\end{center}

\noindent 
The whole string is copied, but in the opposite direction:

\begin{center}
\small
\begin{alltt}
 10101001 10100011 00101010 00001101 11111100
 10110100 11100111 11010011 10111011 00001110
 11000100 00011101 00110010 11111000 11101001
 11110001 00011111 00101110 00111110 10011001
 01110000 01000110 11100001 10111011 10010111
 11001110 01011010 01111111 01100000 10101001
 10001011 0010101.
\end{alltt}
\end{center}

The string whose length is a multiple of 512 is obtained,
by duplicating enough this string and truncating at the next multiple of 512.
This string, in which the whole original text is contained, is denoted by $D$.
Finally, we split our obtained string into blocks of 256 bits and apply
to them the exclusive-or function, from the first two blocks 
to the last one. It results a 256 bits sequence, that is in our example:

\begin{center}
\small
\begin{alltt}
 11111010 11100101 01111110 00010110 00000101
 11011101 00101000 01110100 11001101 00010011
 01001100 00100111 01010111 00001001 00111010
 00010011 00100001 01110010 01000011 10101011
 10010000 11001011 00100010 11001100 10111000
 01010010 11101110 10000001 10100001 11111010
 10011101 01111101.
\end{alltt}
\end{center}
The configuration $x^0$ 
is the result of this pre-treatment  and is a sequence of $n=256$ bits.
Notice that some distinct texts lead to the same string. 

Let us build now the strategy $(S^t)^{t \in \Nats}$ that depends on both 
the original message and a given key.

\subsection{Computing $(S^t)^{t \in \Nats}$}
\label{subset:Computing St}

To obtain the strategy $S$, 
an intermediate sequence $(u^t)^{t \in \Nats}$ is constructed from $D$, as follows:

\begin{enumerate}
\item $D$ is split into blocks of 8 bits. 
Let  $(u^t)^{t \in \Nats}$ be the finite sequence where $u^t$ is 
the decimal value of the $t^{th}$ block.

\item A circular rotation of one bit to the left is applied to $D$ 
  (the first bit of $D$ is put on the end of $D$). 
  Then the new string is split into blocks of 8 bits another time.
  The decimal values of those blocks are added to $(u^t)$.
\item This operation is repeated again 6 times.
\end{enumerate}

Because of the function  $\theta \longmapsto 2\theta ~(mod ~1)$ 
is known to be chaotic in the sense of Devaney~\cite{devaney}, 
we define the strategy $(S^t)^{t \in \Nats}$ with:
$$
S^t=(u^t+2\times S^{t-1}+t) \mod 256,
$$
\noindent which is then highly sensitive to initial conditions and then to 
changes of the original text. 
On the one hand, when a keyed hash function is desired, this sequence $(S^t)^{t \in \Nats}$ is initialized with the given key $k$ (\emph{i.e.}, $S^0=k$).
On the other hand, it is initialized to $u^0$ if the hash function is unkeyed.

\subsection{Computing the digest}
\label{subsec:computing the digest}

To construct the digest, chaotic iterations of $G_{\neg}$ are realized 
with initial state  $X^0=((S^t)^{t \in \Nats},x^0)$ as defined above.
The result of these iterations is a $n=256$ bits vector.
Its components are taken 4 per 4 bits and translated into hexadecimal numbers,
to obtain the hash value:

\begin{center}
\begin{alltt}
    63A88CB6AF0B18E3BE828F9BDA4596A6
    A13DFE38440AB9557DA1C0C6B1EDBDBD.
\end{alltt}
\end{center}

As a comparison if we replace 
\textquotedblleft \textit{%
  The original text}\textquotedblright\ by \textquotedblleft \textit{the original text}\textquotedblright , the hash function returns:

\begin{center}
\begin{alltt}
    33E0DFB5BB1D88C924D2AF80B14FF5A7
    B1A3DEF9D0E831194BD814C8A3B948B3.
\end{alltt}
\end{center}

We then investigate qualitative properties of this algorithm.   


%% file: avalanche.tex
We show in this section that, as a consequence of recalled theoretical
results,  this   hash  function  tends  to   verify  desired  informal
properties of a secure keyed one-way hash function.

\subsection{The avalanche criteria}
\label{subsec:avalanche}

Let  us first  focus  on the  avalanche  criteria, which  means that  a
difference  of  one  bit between  two  given  medias  has to  lead  to
completely different digest.  In our opinion, this criteria is implied
by the  topological properties of sensitive dependence  to the initial
conditions,  expansivity,  and Lyapunov  exponent.  These notions  are
recalled below.

First,  a  function  $f$  has  a  constant  of  expansivity  equal  to
$\varepsilon $ if an arbitrarily  small error on any initial condition
is  \emph{always}  magnified till  $\varepsilon  $.  In our  iteration
context  and  more  formally,  the function  $G_{\neg}$  verifies  the
\emph{expansivity} property if there exists some constant $\varepsilon
>0$ such that for any $X$ and $Y$ in $\mathcal{X}$, $X \neq Y$, we can
find           a            $k\in           \mathbb{N}$           s.t.
$d(G^k_{\neg}(X),G^k_{\neg}(Y))\geqslant \varepsilon$.  We have proven
in~\cite{gfb10:ip}  that,  $(\mathcal{X},G_{\neg})$  is  an  expansive
chaotic system. Its constant of expansivity is equal to 1.


Next,   some  dynamical   systems  are   highly  sensitive   to  small
fluctuations  into   their  initial  conditions.    The  constants  of
sensibility  and   expansivity  have  been   historically  defined  to
illustrate this  fact.  However, in  some cases, these  variations can
become  enormous,  can  grow  in   an  exponential  manner  in  a  few
iterations,  and  neither  sensitivity  nor expansivity  are  able  to
measure such a situation.  This is why Alexander Lyapunov has proposed
a new notion  being able to evaluate the  amplification speed of these
fluctuations we now recall:

\begin{definition}[Lyapunov Exponent]
Let be given an iterative system $x^0 \in \mathcal{X}$ and $x^{t+1} = f(x^t)$. 
Its \emph{Lyapunov exponent} is defined by:
$$\displaystyle{\lim_{t \to +\infty} \dfrac{1}{t} \sum_{i=1}^t \ln \left| ~f'\left(x^{i-1}\right)\right|}$$
\end{definition}

By  using  a topological  semi-conjugation  between $\mathcal{X}$  and
$\mathds{R}$, we  have proven in~\cite{GuyeuxThese10}  that For almost
all $X^0$, the Lyapunov exponent of chaotic iterations $G_{\neg}$ with
$X^0$ as initial condition is equal to $\ln (n)$.

Let us now explain why the topological properties of our hash function
lead to the  avalanche effect. Due to the  sensitive dependence to the
initial condition, two close  media can possibly lead to significantly
different  digests.   The  expansivity  property  implies  that  these
similar medias mostly lead to  very different hash values.  Finally, a
Lyapunov exponent greater than 1 lead to the fact that these two close
media will always finish to have very different digests.

%% file: preimage.tex
\subsection{Preimage Resistance}

Let us now discuss about the first preimage resistance of our unkeyed hash function denoted by $h$.
Indeed, as recalled previously, an adversary given a target image $D$ should not be able to find a preimage $M$ such that $h(M)=D$.
One reason (among many) why this property is important is that on most computer systems user passwords are stored as the cryptographic hash of the password instead of just the plaintext password.
Thus an attacker who gains access to the password file cannot use it to then gain access to the system, unless it is able to invert target message digest of the hash function.

We now explain why, topologically speaking, our hash function is resistant to preimage attacks.
Let $m$ be the message to hash, $(S,x^0)$ its normalized version (\emph{i.e.}, the initial state of our chaotic iterations), and $M=h(m)$ the digest of $m$ by using our method.
So chaotic iterations with initial condition $(S,M)$ and iterate function $G_{\neg}$ have $x^0$ as final state.
Thus it is impossible to invert the hash process with a view to obtain the normalized message by using the digest.
Such an attempt is equivalent to try to forecast the future evolution of chaotic iterations by only using a partial knowledge of its initial condition.
Indeed, as $M$ is known but not $S$, the attacker has an incertitude on the initial condition.
he only knows that this value is into an open ball of radius 1 centered at the point $M$, and the number of terms of such a ball is infinite.

With such an incertitude on the initial condition, and due to the numerous chaos properties possessed by the chaotic iterations (as these stated in Section~\ref{subsec:avalanche}), this prediction is impossible.
Furthermore, due to the transitivity property, it is possible to reach all of the normalized medias, when starting to iterate into this open ball.
Indeed, it is possible to establish that, all of these possible normalized medias can be obtained in at most 256 iterations, and we iterate at least 519 times to obtain our hash value (\emph{c.f.} Proposition~\ref{prop:nombre d'itérations} below). Finally, to find the normalized media does not imply the discovery of the original plain-text.

%% file: confusion.tex

%% file: performance.tex
Let us first give some examples of hash values
before discussing about the algorithm complexity.


\subsection{Hash values}

\begin{figure}
\centering
\includegraphics[width=6cm]{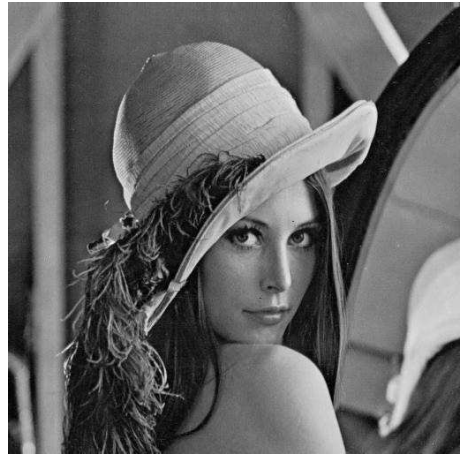}
\caption{The original plain-image.}
\label{plain-image}
\end{figure}

Let  us  now  consider  our  hash  function  with  $n=128$.   To  give
illustration of  the confusion and  diffusion properties, we  will use
this function to generate hash values in the following cases:
\begin{description}
\item[Case 1.]~ The original text message is the poem \textit{Ulalume}
(E.A.Poe),  which is  constituted by  104 lines,  667 words,  and 3754
characters.
\item[Case 2.]~ We change  \textit{serious} by \textit{nervous} in the
verse ``\textit{Our talk had been serious and sober}''
\item[Case 3.]~ We replace the last point `.' with a coma `,'.
\item[Case 4.]~  In ``\textit{The skies they were  ashen and sober}'',
skies becomes Skies.
\item[Case  5.]~ The  new original  text is  the binary  value  of the
Figure~\ref{plain-image}.
\item[Case 6.]~  We add 1  to the gray  value of the pixel  located in
position (123,27).
\item[Case 7.]~ We substract 1 to  the gray value of the pixel located
in position (23,127).
\end{description}

The corresponding hash values in hexadecimal format are:

\begin{description}
\item[Case 1.]~\textsf{C0EA2325BBF956D27C3561977E48B3E1},
\item[Case 2.]~\textsf{6C4AC2F8579BCAB95BAD68468ED102D6},
\item[Case 3.]~\textsf{A538A76E6E38905DA0D35057F1DC1B14},
\item[Case 4.]~\textsf{01530A057B6A994FBD3887AF240F849E},
\item[Case 5.]~\textsf{DE188603CFE139864092C7ABCD21AE50},
\item[Case 6.]~\textsf{FF855E5A626532A4AED99BACECC498B1},
\item[Case 7.]~\textsf{65DB95737EFA994DF37C7A6F420E3D07}.
\end{description}

 These simulation results are coherent with the topological properties
of  sensitive dependence  to the  initial condition,  expansivity, and
Lyapunov exponent: any alteration  in the message causes a substantial
difference in the final hash value.

\subsection{Algorithm Complexity}

In this section is evaluated the complexity of the above hash  function for a size $l$ of the media (in bits).

\begin{proposition}
The stages of initialization 
(Sections~\ref{subsec:computing x0} and~\ref{subset:Computing St}) 
need $\mathcal{O}(l)$ elementary operations to be achieved.
\end{proposition}
\begin{IEEEproof}
In this stage only linear operations over binary tables are achieved, such as: copy, circular shift, or inversion.
\end{IEEEproof}
Let us consider the digest computation stage (Section~\ref{subsec:computing the digest}).

\begin{proposition}
\label{prop:nombre d'itérations}
The digest computation stage requires less
than $2l + 2 \log_2(l+1) +515$ elementary operations.
\end{proposition}

\begin{IEEEproof} The cost of an  iteration is reduced to the negation
operation  on a  bit, which  is  an elementary  operation.  Thus,  the
second stage  is realized in  $t$ elementary operations, where  $t$ is
the  number of  terms into  the sequence  $S$.  But  $S$ has  the same
number of terms than $u$, and  $u$ and $D$ have the same size (indeed,
to build $u$,  $D$ has been copied 8 times, and  bits of this sequence
have been regrouped 8  per 8 to obtain the terms of  $u$).  To sum up,
the size of $D$ is equal  to the total number of elementary operations
of the digest computation stage.

The following operations are realized to obtain $D$.
\begin{enumerate}
\item The digit 1 is added: $D$ has $l + 1$ bits.
\item The binary value of the size is added, followed by another bit: $D$ has \linebreak $l+2 + \log_2(l+1)$ bits.
\item This string is copied after inversion: $D$ has now \linebreak $2 \times \left( l+2 + \log_2(l+1) \right)$ bits.
\item Lastly, this string is copied until the next multiple of 512: in the worst situation, 511 bits have been added, so $D$ has in the worst situation $2l + 2 \log_2(l+1) +515$ bits.
\end{enumerate}
\end{IEEEproof}

We can thus conclude that:

\begin{theorem}
The computation of an hash value is linear with the hash function presented in this research work.
\end{theorem}

%% file: perf.tex
\subsection{Experimental Evaluation}

We  focus now  on  the  illustration of  the  diffusion and  confusion
properties~\cite{Shannon49}.  Let  us recall that  confusion refers to
the desire to make the relationship between the key and the ciphertext
as complex and involved as  possible, whereas diffusion means that the
redundancy in the statistics of  the plaintext must be "dissipated" in
the statistics of the  ciphertext.  Indeed, the avalanche criterion is
a modern  form of the  diffusion, as this  term means that  the output
bits should depend on the input bits in a very complex way.

\begin{figure}[t]
\centering
\subfloat[Original text (ASCII)]{
\includegraphics[scale=0.29]{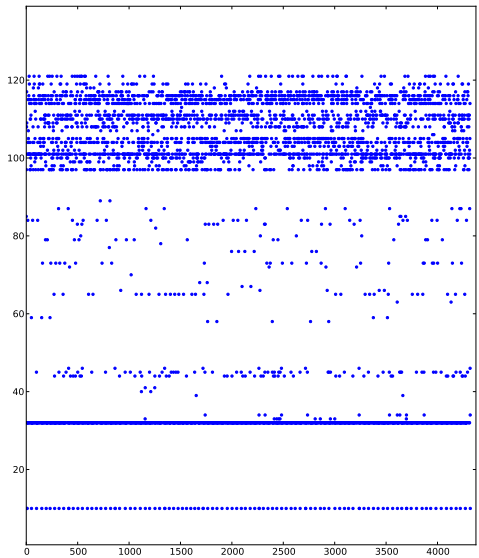}
\label{fig:ASCII repartition0}
} 

\subfloat[Digest (Hexadecimal)]{
\includegraphics[scale=0.25]{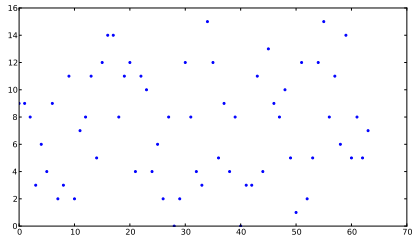}
\label{fig:Hexa repartition0}
}
\caption{Values repartition of Ulalume poem}
\end{figure}

\subsubsection{Uniform repartition for hash values}

To show the diffusion and confusion properties verified by our scheme,
we first  give  an  illustration  of  the  difference  of  characters
repartition between a plain-text and  its hash value when the original
message is again the Ulalume poem.
In Figure~\ref{fig:ASCII  repartition0}, the ASCII codes 
are localized within a small area, 
whereas in Figure~\ref{fig:Hexa repartition0}  the hexadecimal numbers of 
the hash value are uniformly distributed.

\begin{figure}[t]
\centering
\subfloat[Original text (ASCII)]{
\includegraphics[scale=0.15]{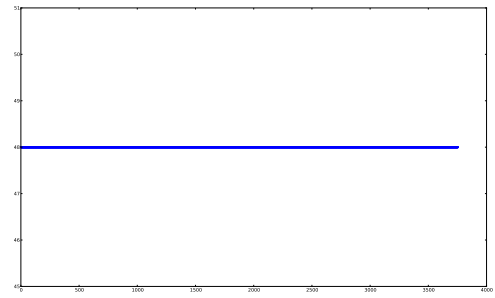}
\label{fig:ASCII repartition}
}

\subfloat[Digest (Hexadecimal)]{
\includegraphics[scale=0.15]{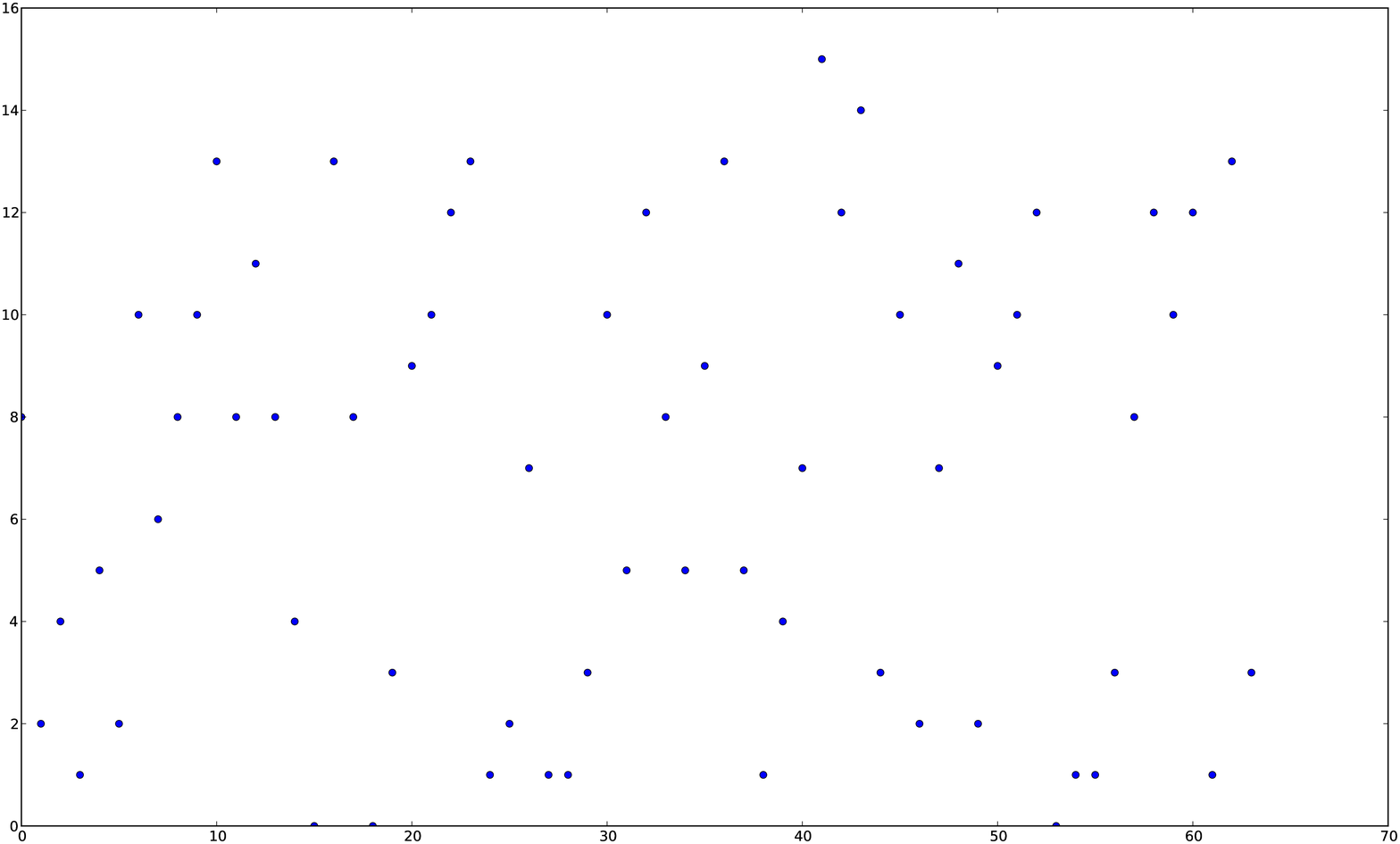}
\label{fig:Hexa repartition}
}
\caption{Values repartition of the ``\textit{00000000}'' message}
\end{figure}

A similar experiment has been realized with a message having the same size,
but which is only constituted by the character ``\textit{0}''. 
The contrast between the plain-text message and its digest 
are respectively presented in Figures~\ref{fig:ASCII repartition} 
and~\ref{fig:Hexa repartition}.
Even under this very extreme condition, the distribution of the digest still remains uniform. 
To conclude, these simulations tend to indicate that no information concerning the original message can be found into its hash value, as it is recommended by the Shannon's diffusion and confusion.

\subsubsection{Behavior through small random changes}

\begin{figure}
\centering
\includegraphics[scale=0.7]{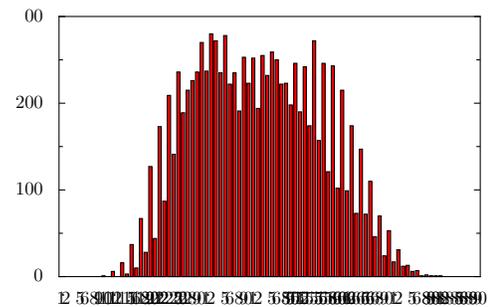}
\caption{Histogram}
\label{Histogram}
\end{figure}

We now consider the following experiment.
A first message of 100 bits is randomly generated, and its hash value of size 80 bits is computed.
Then one bit is randomly toggled into this message and the digest of the new message is obtained.
These two hash values are compared by using the hamming distance, to compute the number $B_i$ of changed bits.
This test is reproduced 10000 times.
The corresponding distribution of $B_i$ is presented in Figure~\ref{Histogram}.

As desired, Figure~\ref{Histogram} show that the distribution is centered around 40, which reinforces the confidence put into the good capabilities of diffusion and confusion of the proposed hash algorithm.

\subsubsection{Statistic analysis of diffusion and confusion}

Finally, we generate 1000 sequences of 1000 bits, and for each of these sequences, we toggle one bit, thus obtaining a sequence of 1000 couples of 1000 bits.
As previously, the two digests of each couple $i$ are obtained, and the hamming distance $B_i$ between these digests are computed.
To analyse these results, the following common statistics are used.
\begin{itemize}
\item Mean changed bit number $\overline{B} = \frac{1}{N}\sum_{i=1}^{\mathsf{N}} B_i$.
\item Mean changed probability $P = \frac{\overline{B}}{128}$.
\item $\Delta B = \sqrt{\dfrac{1}{N-1}\sum_{i=1}^\mathsf{N} (B_i-\overline{B})^2}$.
\item $\Delta P = \sqrt{\dfrac{1}{N-1}\sum_{i=1}^\mathsf{N} (\frac{B_i}{128}-P)^2}$.
\end{itemize}

The obtained statistics are listed in Table~\ref{table:statistical performances}. 
Obviously, both the mean changed bit number $\overline{B}$ and the mean changed probability $P$ are close to the ideal values (64 bits and 50\%, respectively), which illustrates the diffusion and confusion capability of our algorithm. 
Lastly, as $\Delta B$ and $\Delta P$ are very small, these capabilities are very stable.

\begin{table}
\begin{tabular}{ccccccc}
\hline
 & $B_{min}$ & $B_{max}$ & $\overline{B}$ & $P(\%)$ & $\Delta B$ & $\Delta P(\%)$ \\
\hline
N = 256  & 50 & 92 & 67.57 & 52.78 & 8.89 & 6.95  \\
N = 512  & 47 & 82 & 65.13 & 51.11 & 7.65 & 5.87  \\
N = 1024 & 47 & 81 & 63.01 & 52.10 & 7.51 & 5.71  \\
\hline
\end{tabular}
\caption{Statistical performance of the proposed hash function}
\label{table:statistical performances}
\end{table}
\enlargethispage{-0.4in}

%% file: conclusion.tex
MD5 and SHA-0 have been broken in 2004. 
An attack over SHA-1 has been achieved with only $2^{69}$ operations (CRYPTO-2005), that is, 2000 times faster than a brute force attack (that requires $2^{80}$ operations).
Even if $2^{69}$ operations still remains impossible to realize on common computers, such a result based on a previous attack on SHA-0 is a very important one:
it leads to the conclusion that SHA-2 is not as secure as it is required for the Internet applications.
So new original hash functions must be found.

In this research work, a new hash function has been presented.
The security in this case has been guaranteed by the unpredictability of the behavior of the proposed algorithms.
 The algorithms derived from our approach satisfy important properties of topological chaos such as sensitivity to initial conditions, uniform repartition (as a result of the transitivity), unpredictability, and expansivity.
 Moreover, its Lyapunov exponent can be as great as needed.
The results expected in our study have been experimentally checked. The choices made in this first study are simple: compression function inspired by SHA-1, negation function for the iteration function, \emph{etc.} 
The aim was not to find the best hash function, but to give simple illustrated examples to prove the feasibility in using the new kind of chaotic algorithms in computer science.
Finally, we have shown how the mathematical framework of topological chaos offers interesting qualitative and qualitative tools to study the algorithms based on our approach.

In future work, we will investigate other choices of iteration functions and chaotic strategies. We will try to characterize topologically the diffusion and confusion capabilities.  Other properties induced by topological chaos will be explored and their interest for the realization of hash functions will be deepened.